\documentstyle[12pt]{article}
\begin{document}
\title{Heavy quarkonia}
\author{Kacper Zalewski\thanks{Supported in part by the
KBN grant 2P302-076-67}\\
Jagellonian University\\ and\\ Institute of Nuclear Physics, Krak\'ow, Poland}
\maketitle

\begin{abstract}
Two complementary approaches to the theory of heavy quarkonia are discussed.
The nonrelativistic potential models give amazingly accurate predictions, but
lack a theoretical justification. The expansion in powers of $v/c$ is
theoretically very acceptable, but is not as good in giving numerical
predictions. The importance of combining these two approaches is stressed.
\end{abstract}

\section{INTRODUCTION}

The subject of this presentation are bound states in the $\overline{b}b$,
$\overline{b}c(\overline{c}b)$ and $\overline{c}c$ systems. In the old days
it was usual to write that the heavy quarkonia containing the $t$-quark, the
$\overline{t}$-antiquark or both will be the most interesting ones . Now,
however, it is known that the lifetime of the $t$-(anti)quark is too short for
hadronization, so that the quarkonia containing $t$-(anti)quarks do not exist.
We limit our discussion to bound states below the thresholds for strong decays,
i.e. below $2M_B = 10.558$ GeV for the $\overline{b}b$ system, below $M_B + M_D
= 7.146$ GeV for the $\overline{b}c\;(\overline{c}b)$ system and below $2M_D =
3.690$ GeV for the $\overline{c}c$ system. We also ignore purely relativistic
effects like hyperfine splittings, fine splittings etc.

Plotting the excitation energies $M - M(^3S_1)$ for the three families of
quarkonia one makes the following two observations. Firstly, the spectrum of
excitations in the range below the thresholds for strong decays ($M_{th}$)
depends little on the quark masses. Secondly, $M_{th} - M_{{}^3S_1}$ increases
with the quark masses. Thus the number of bound states below $M_{th}$,
ignoring the fine and hyperfine splittings, is three for $\overline{c}c$, four
for $\overline{b}c(\overline{c}b)$ and ten for $\overline{b}b$. Actually,
many of these states have not yet been observed, but the predictions of the
nonrelativistic potential models are so reliable  that there is little doubt
about the correctness of this counting.

We shall discuss two complementary approaches. Nonrelativistic potential models
give for many quantities predictions in excellent agreement with experiment. On
the othere hand their relation to accepted theory is unknown and they have
serious consistency problems. For this reason they are often interpreted as
recipes rather than as theoretical models. Nevertheless, they are generally
used, when realistic predicitons are necessary. The recently proposed method
of expanding in powers of $v/c$ (the heavy quarks are slow) \cite{BBL},
\cite{MSC}, \cite{KHO} has much better theoretical foundations. On the other
hand its predictive power cannot (yet?) compete with that of the
nonrelativistic potential models in the range of their applicability.
Nevertheless, this approach has produced some general results of great
interest. We shall show how it resolved a problem, which has resisted attempts
to clarify it for almost twenty years.

\section{NONRELATIVISTIC POTENTIAL MODELS}

There are many nonrelativistic potential models for heavy quarkonia and each of
them has been introduced for some good reason. Thus e.g., among the more than
ten published models using potentials of the form

\begin{equation}
V(r) = ar^\alpha - br^{-\beta} + c,
\end{equation}
where the constants $a,b,\alpha,\beta$ are nonnegative, the Cornell potential
$\alpha = \beta = 1$ \cite{EIC} becomes Coulombic for $r \rightarrow 0$ and
stringy for $r \rightarrow \infty$ as expected from QCD; the logarithmic
potential of Quigg and Rosner $\alpha = \beta \rightarrow 0$ \cite{QRO} gives
an excitation spectrum, which does not depend on the quark masses; Martin's
potential $\alpha = 0,\;\beta = 0.1$ \cite{MAR} makes it possible to perform a
particularly elegant mathematical analysis. Since the point we want to stress
is the predictive power of the nonrelativistic potential models, we choose the
potential \cite{MZA}

\begin{equation}
V(r) = 0.706380\left[\sqrt{r} - \frac{0.460442}{r}\right] + 8.81715,
\end{equation}
where $r^{-1}$ and $V(r)$ are in units of GeV. The corresponding mass of the
$b$ quark is $m_b = 4.80303$ GeV. This potential and this quark mass have been
chosen so as to optimize the agreement of the predictions with the following
measured quantities characterizing the $\overline{b}b$ system: the masses
$M(1S),\;M(1P)$, $M(2S),\;M(2P),\;M(3S)$; the squared moduli of the wave
functions at zero separation between the quark and the antiquark
$|\psi_{1S}(0)|^2,\; |\psi_{2S}(0)|^2,\;|\psi_{3S}(0)|^2$; the absolute values
of the dipole electric transition matrix elements  $|\langle 1P|r|2S\rangle|$,
 $|\langle 2P|r|3S\rangle|$ and the ratio of such absolute values $
|\langle 1S|r|2P\rangle|/|\langle 2S|r|2P\rangle|$. Actually, instead of
fitting the eleven observables with four free parameters, we fitted \cite{MZA}
eight observables with one partameter and fixed the other three parameters so
as to reproduce exactly the remaining three observables. This has little effect
on the quality of the fit and makes the analysis simpler and more meaningful
statistically. Another advantage of this procedure is that the poorly known
corrections in the evaluation of the squares of the wave functions at the
origin from the directly measured leptonic decay widths almost drop out.

For the parameters quoted above one finds $\chi^2 = 6.5$ for seven degrees of
freedom, which means an excellent fit. This is nontrivial, because the accuracy
of the experimental data is high. In particular, the four masses are measured
with uncertainties ranging from $0.2$ MeV to $0.5$ MeV, i.e. with an error
below 0.01\%!. Incidentally, our parmeters are given with six digit accuracy,
only in order to enable the reader to check our computer calculations.

One of the early successes of the nonrelativistic quark models was that the
potentials tuned to describe the $\overline{c}c$ quarkonia have also given an
acceptable description of the $\overline{b}b$ quarkonia, when these were
discovered several years later. This fact, which had been expected in the
nonrelativistic quark models "QCD is flavour blind", is not predicted any more
in the modern approach based on the expansion in powers of velocity. Therefore,
it is of interest to check, how this universality is satisfied with the present
data. Introducing one more parameter $m_c = 1.3959$ GeV and assuming that the
constant in the potential changes by $2(m_c - m_b)$, we have calculated for the
$\overline{c}c$ quarkonia the values of the observables
$M(1S),\;M(1P),\;M(2S),\;|\psi_{1S}(0)|^2,\;|\psi_{2S}(0)|^2$ $|\langle
1S|r|1P\rangle|$ and $|\langle 1P|r|2S\rangle|$. The masses agree with
experiment within $4$ MeV. This margin is about an order of magnitude more than
the experimental uncertainties, but on the absolute scale it is not much. The
wave functions and the electric dipole matrix elements have to be rescaled (a
rescaling factor of $1.3$ for the wave func\-tions and a rescaling factor of
0.73 for the matrix elements), but then they agree with ex\-periment. Some
rescaling is expected, because when extracting the quantities to be compared
with the model from the directly measurable ones we used the same correction
factors as for the $\overline{b}b$ system, which in the case of the wave
functions is certainly, and in the case of the matix elements very probably, a
rather crude approximation only. We conclude that deviations from the
$\overline{b}b$ --- $\overline{c}c$ universality are clearly seen, but that
they are not very large.

Let us conclude by stressing that the nonrelativistic potential model cannot be
interpreted na\"{\i}vely as a consistent theory. E.g. within this model one can
calculate the kinetic energies of the quarks and from that their mean square
velocities. The results are $\langle v^2\rangle \approx 0.25 c^2$ for the
ground state of the $\overline{c}c$ system and $\langle v^2\rangle \approx 0.08
c^2$ for the ground state of the $\overline{b}b$ system. This is inconsistent
with the assumption that high precision results can be derived assuming that
the motion is nonrelativistic. To be sure, one can speculate that relativistic
corrections can be absorbed into redefinitions of the nonrelativistic
parameters of the model, but before this is demonstrated, the main argument in
favour of the nonrelativistic potential models remains their amazing success in
predicting experimental results.

Let us consider now a more formal approach.

\section{EXPANSION IN POWERS OF VELOCITY}

Since confinement is a nonperturbative effect, there is little hope of
obtaining a good description of heavy quarkonia by summing Feynman diagrams.
Another obvious idea -- to try an expansion in inverse powers of the heavy
quark mass -- is also unlikely to work. Arguments derived from quantum field
theory can be found in ref \cite{MSC}. Here we shall qualitatively discuss the
high  mass limit using the Cornell potential

\begin{equation} V(r) = ar - \frac{b}{r} + c, \end{equation} where $a>0,\;b>
0$. It is plausible that in the high mass limit the system becomes Coulombic. A
Coulombic system with particle masses $M$ has a radius of order $\frac{1}{M}$,
so that in the high mass limit it is consistent to neglect the $ar$ term. Let
us try, however, to consider this term as a perturbation. For $a<0$ the
potential is unbounded from below. There is no ground state and the
perturbation series cannot converge. Theorems about the convergence radii of
power series guarantee that if the perturbation series in $a$ is divergent for
some $a_0<0$, it will also diverge for all positive $a$ from the range
$0<a<-a_0$. Thus $a=0$ is a singular point. This could be harmless. E.g. the
perturbative expansion in QCD is believed to be an (asymptotic) power series
expansion around a singular point. The reason for its success is that the first
terms of this expansion approach the correct result so rapidly that they are
enough for most practical applications and, therefore, the convergence or
divergence of the whole series is irrelevant. In the quarkonium problem,
however, the high mass limit -- with the wave function localized in an
infinitesimal region around $r=0$ -- is so remote from any plausible
description of a quarkonium that there is little hope that the first few terms
of the expansion will make it realistic. Thus, one has to look for another
idea.

Let us note an important implication \cite{BBL} of the observation that in the
high mass limit the quarkonium is approximately a Coulombic system. Since the
kinetic and the potential energies should be comparable, we expect for high
masses

\begin{equation}
\frac{\alpha_s(\frac{1}{R})}{R} \approx M v^2,
\end{equation}
where $R = \frac{1}{Mv}$ is an estimate of the quarkonium radius. Since
$\alpha_s(\mu)$ is a decreasing function of its argument and since $v$ (which
is in units of the velocity of light $c$) is less than one, this implies

\begin{equation}
v > \alpha_s(M).
\end{equation}
Thus, it is inconsistent to include radiative corrections of order
$O(\alpha_s^k)$ without also including the relativistic corrections of order
at least $v^k$.

The new and promising idea \cite{BBL},\cite{MSC},\cite{KHO} is to combine
factorization and an expansion in powers of the quark velocity (in the rest
frame of the quarkonium and in units of $c$) $v$. Thus, any probability
amplitude, e.g. the probability amplitude for the decay of quarkonium
$\overline{Q}Q$ into a gluon pair, is represented as a sum of terms. Each of
these terms is a product of a soft matrix element, which in principle is
obtainable from a lattice calculation, but at present is not known, and of a
hard matrix element, which can be calculated perturbatively. Using suitable
scaling rules it is possible to ascribe to each term an order $n$, which means
that for $v\rightarrow 0$ this term is of order $O(v^n)$. At each order there
is a finite number of terms only. The leading term approximation consists of
all the terms of lowest order. It may be systematically improved by including
higher and higher order terms.

In this approach the soft matrix element depends on the quark mass in a way,
which is beyond our control. Thus, the simple universality from the
nonrelativistic quark model is lost. Nevertheless, many approximations known
from other approaches, like spin symmetry, vacuum saturation or some of the
relations between the production and decay amplitudes can in many cases be
rigorously justified at sufficiently low orders of the expansion in powers of
$v$. Moreover, the results have often simple physical interpretations.

Perhaps the most impressive success of this approach is the analysis of the
decays of $P$-wave quarkonia into light hadrons \cite{BBL1}, \cite{BBL}. This
process had been studied for a long time in the framework of the
nonrelativistic potential models. One finds that, since for $P$-states
$|\psi(0)|^2 = 0$, $Q$ and $\overline{Q}$ have small probability to get so
close to each other as to be able to annihilate with a significant
probability. As a result, the decay probability amplitude is reduced (compared
to the decay probabilities of the $S$-states) by a factor of order $O(v)$. The
problem is, however, that this comparatively small amplitude has infinite QCD
corrections! The infrared divergences in the calculation do not cancel
\cite{BGR}.

In the spirit of the expansion in powers of $v$, we must look for other
contributions to the decay amplitude, which are of the same order in $v$. There
is one more such term, where the quarkonium component consists of a
$\overline{Q}Q$ system in a colour octet $S$-state accompanied by a "dynamical"
gluon so that the whole quarkonium is a colour singlet as it should. This
component is small -- of order $O(v)$ as compared to the main term. Its
$\overline{Q}Q$ part, however, being an $S$-state annihilates easily, so that
the contribution of this component to the decay amplitude of the quarkonium
into light hadrons is of the same order in $v$ as the decay amplitude of the
$P$-wave term. Thus, both terms must be included in a correct leading order
calculation. After this is done, the infrared singularities in the QCD
corrections cancel \cite{BBL1} and a finite, acceptable result is obtained.
Thus, at least in principle, the problem posed in 1976 has been solved.

\section{CONCLUSIONS}

The expansion in powers of the velocity $v$ is a respectable physical approach.
It can be used to prove statments, which in the nonrelativistic potential
approach had to be, often implcitly, conjectured. It also introduces important
corrections beyond the scope of the potential approach, like the contribution
of the $|(\overline{Q}Q),\mbox{gluon}\rangle$ state to the annihilation of the
$P$-wave quarkonia into light hadrons. Nevertheless, for making practically
useful predictions it cannot, at present, replace the potential models. The key
problem seems to be, how to combine the advantages of the two approaches? How
to calculate the soft matrix elements of the expansion in powers of the
velocity $v$ from potential models, or how to justify the potential models
using the powerful formalism of the expansion in powers of $v$?

\end{document}